# STRUCTURE, VELOCITY FIELD AND
# TURBULENCE IN NGC 604

Medina Tanco G. A.[1,2], Sabalisck N. [1]

Jatenco-Pereira V. [1], Opher R[1]

[1]Instituto Astronômico e Geofísico, USP, Av. Miguel Stéfano 4200, CEP 04301-904, São Paulo, SP, BRASIL

[2] Royal Greenwich Observatory, Madingley Rd., Cambridge, CB3 0EZ, UK

(e-mail: gmt@ast.cam.ac.uk)

## Abstract

The Hα peak intensity, velocity shift and velocity dispersion maps of the giant HII region NGC 604 in M 33, obtained by two dimensional high spatial resolution Fabry-Perot observations with TAURUS II at the 4.2 m William Herschel Telescope in Spain (Sabalisck, 1995), are analyzed via two point correlation functions. The whole system seems to rotate as a rigid body on scales from 50 to 80 pc (the largest studied scale), with a period of ≈ 85 Myr. We demonstrate that the cloud seems to be comprised of eddies with varying characteristic scale lengths which range from 10 pc to the largest observed scales. The calculated kinetic energy spectrum may be interpreted as either a manifestation of a double cascading spectrum of forced two-dimensional turbulence, or as a Kolmogorov three-dimensional turbulence (although this last possibility seems unlikely). According to the first interpretation, turbulence is being forced at scales of ≈ 10 pc, while an enstrophy (mean-square vorticity) cascade has developed down to the smallest scales resolved and an inverse kinetic energy cascade extends up to scales of ≈ 70 pc where a low wave number turn over is observed; if true, this would be the first time that such a phenomenon has been observed outside the Solar System. As for the second interpretation, energy should be injected at the largest scales, ≈ 70 pc. In both cases the average intrinsic optical depth consistent with the results is ≈ *20* pc.

# 1. Introduction

The central part of the giant, high surface brightness, HII region NGC 604 (see, Figure 1.a) located in M33, a galaxy in the Local Group, shows several emitting knots, and a large inhomogeneous halo where several shells, filaments and loops are distributed (D'Odorico and Rosa 1981).

NGC 604 is known to be a very young star-forming region, $\approx 4 \times 10^6$ yr old (D'Odorico and Rosa 1981). The HI mass in the region is $\approx 1.6 \times 10^6$ $M_\odot$ ( Wright 1971), whereas the ionized mass is $\approx 10^6 M_\odot$ (Israel et al. 1982), and its integrated H$\alpha$ luminosity is $\approx 4.5 \times 10^{39}$ ergs sec$^{-1}$ (Melnick 1992).

Recently, the H$\alpha$ intensity peak, velocity shift and velocity dispersion maps of NGC 604, were obtained by means of two dimensional high spatial resolution Fabry-Perot observations with TAURUS II at the 4.2 m William Herschel Telescope in Spain. The kinematics of this region was studied in order to show the behavior of the gas at different scales of motion (Sabalisck et al. 1995).

The turbulent movement inside the cloud has been studied by different authors (e.g., Smith and Weedman 1970; Melnick 1977, 1980; Terlevich and Melnick 1981; D'Odorico and Rosa 1981; Rosa and D'Odorico 1982; Rosa and Solf 1984; Hippelein and Fried 1984; Skillman and Balick 1984; Hunter and Gallagher 1985a,b; Roy et al. 1986; etc.). The behavior of the turbulence at large scales is different from that at small scales, which led to different interpretations of the mechanisms at play inside the region covering autogravitation, stellar winds and the Champagne effect. Nevertheless, the models proposed hitherto to explain the kinematics of NGC 604, based mainly on line profile analysis, are not conclusive.

In the present study, we are also concerned with the velocity structure of the cloud and the character of the observed turbulence. We study the auto-correlation and second order structure functions of the components of the velocity field parallel to the line of sight. We demonstrate that the cloud presents a rich hierarchical velocity structure, and is  made up of fragments with various characteristic scale lengths, the most notorious  having scales of $\approx$ *10* pc, while the largest may be a billow-like structure comprising the region as a whole and rotating with a period of $\approx$ *85* Myr.

Besides, we find strong evidences for a Kolmogorov like spectrum at scales ≥ 10 pc, which changes rapidly to another very well defined power law at smaller scales. The existence of both power laws may be understood as either a two- or a three-dimensional turbulent spectrum modified by optical depth effects.

If turbulence were, in fact, two-dimensional, then an inverse energy cascade is taking place inside the cloud. The source of energy should come from scales of ≈ *8-10* pc. A possible source of this energy is the expansion of the terminal shocks of the wind bubbles of individual OB stars or OB associations, or the tens of WR stars known to exist in the region. This would be, to our knowledge, the first time that two-dimensional turbulence is observed beyond the Solar System (see, for example, Gierasch (1996) and Cho and Polvani (1996), for application to the upper atmospheres of the giant outer planets).

In section 2 we present the data used in the analysis, and in section 3 we discuss the velocity structure of the cloud and turbulence observed. Our conclusions are summarized in section 4.

## 2. The data

The two-dimensional maps of the HII region were constructed using a high resolution aperture equivalent to *0.26* x *0.26* arcsec². The emission spectrum obtained through this area was adjusted with a Gaussian fit. From this Gaussian curve, peak intensity, velocity shift (radial velocity) and velocity dispersion values were obtained. The object was scanned with this window and a triad of parameters was obtained for each pixel. Three maps were assembled in this way from a total of ≈ *10 ⁴* spectra (see, Sabalisck, 1995 for further details).

The corresponding peak intensity, and radial velocity maps can be seen in Figures 1a-b. The dimension of the images are 256 x 256 pixels, where each pixel spans over nearly 0.9 pc x 0.9 pc for an assumed distance of 720 Mpc (D'Odorico and Rosa 1981).

## 3. Structure of the cloud and turbulence

The diagnostic tools used here are the auto-correlation and second order structure functions respectively (e.g., Lesieur 1987):

$$F_1 (\underline{r}) = \langle v(\underline{x}) \cdot v(\underline{x} + \underline{r}) \rangle , \tag{1}$$

$$F_2 (\underline{r}) = \langle [ v(\underline{x}) - v(\underline{x} + \underline{r}) ]^2 \rangle , \tag{2}$$

averaged over the velocity image of the cloud. In the previous equations, $v(\underline{x})$ is the peculiar velocity component of the emitting material in a given pixel projected onto the line of sight, i.e., the velocity actually measured for the pixel from the Hα Doppler shift minus the average velocity of the image relative to the observer ( $v(\underline{x}) = u(\underline{x}) - <u(\underline{x}) >$ ). This is the velocity displayed in Figure 1.b and used in the text.

We are concerned here with the dynamic properties of the cloud on the resolved scales, r > 3.6 pc for an assumed distance of 720 Mpc (D'Odorico and Rosa 1981). Therefore, a single Gaussian fit to the line profile is used in each pixel, despite that more than one component exist in many cases as shown by Muñoz-Tuñon et al (1996) and Yang et al (1996). Consequently, we assume that $v(\underline{x})$ is representative of the average velocity of the column of emitting gas inside the pixel, and analyze the validity of this assumption a posteriori.

$F_1 (\underline{r})$ and $F_2 (\underline{r})$ give statistical information averaged over the image. Consider, for example, an eddy with typical rotational velocity $v_r$ and radius $r$. The inertial time, or *turn over* time, of this eddy is ≈ $r/v_r$. If one assumes that this eddy loses an appreciable part of its energy during a turn over time, the energy dissipation rate ε is proportional to $v_r^2/(r/v_r)$. therefore:

$$v_r \approx (\varepsilon \cdot r)^{1/3} . \tag{3}$$

Let us associate a wave number $k = 2\pi/r$ to $r$. The kinetic energy of eddies in a spectral vicinity of k is $\propto \int_{k/10}^{k} E(p)\,dp$ , and, if $E(k)$ decreases following a power law, it is $\propto k\,E(k)$. Then $v_r{}^2 \propto k\,E(k)$ and the previous eq. is equivalent to a Kolmogorov's (Kolmogorov 1941) law:

$$E(k) \propto \varepsilon^{2/3} \cdot k^{-5/3} \tag{4}$$

Note that $v_r$ is a typical velocity difference between two points whose distance $r$ corresponds to the inertial range eddies. In particular, for the second order structure function, $F_2\,(\Delta r)$, we have:

$$F_2\,(\Delta r) \approx (\varepsilon \cdot r)^{2/3}, \tag{5}$$

in the inertial range $l_d < r < l$. Therefore, the structure function gives direct information about the kind of turbulence present in the cloud.

Figure 2 shows the velocity correlation function (1) evaluated over the ninth central portion of the velocity image, Figure 1.b (x ∈ (86,172) and y ∈ (86,172) in pixel coordinates, and labeled as frame "0" in Figure 1.a). This region was chosen to avoid the noise introduced by the irregular borders of the cloud and the lack of measurements outside it. In the case of a random velocity field, the correlation function, as defined by equation (1), presents a maximum for $r=0$ and decays rapidly with increasing $r$. This is clearly not the case in Figure 2, where a high degree of structuring can be seen. In particular, the straight line (a) in this figure shows that the correlation function varies linearly with $r$ for scales in the approximate range $50 < r < 80$ pc, which can be interpreted as rigid body rotation of the cloud on those scales. This seems to be in agreement with a visual inspection of image 1.b, where a long, curved 3D-billow can be visualized. A simple calculation gives a rotation period of $85$ Myr for the feature, something of the order of the lifetime of the cloud. Our numerical simulations show that the rigid rotation of a body with the size of the region considered, would produce similar observational results (from the point of view of the correlation function) with both smaller and larger scales departing from the linear behavior due to border effects. (Consequently, everything above $r \approx 80$ pc must be discarded as noise.)

Similarly, by subtracting the previous linear contribution (a) (thin curve in Figure 2) to the correlation function, it is apparent that another component can exist, superimposed on the former one, which corresponds to rotation on scales $r < 40$ pc.

A secondary maximum of the correlation function is visible at scales of $\approx 20$ pc, which can be interpreted as a manifestation of a spatially periodic velocity field of period $r \approx 20$ pc or, moreover, as volume elements with transverse dimensions of the order of $\Delta r \approx 10$ pc and rotational movement. In order to illustrate the latter behavior, the insert in Figure 2 shows the correlation function after subtracting the hypothetical rotation component on scales $r < 40$ pc, straight line (b).,

Therefore, the correlation function gives indications of the possible existence of a system of eddies spanning various scales ($\approx 80$, $40$, $20$ and $10$ pc) and extending over the whole region.

Further information can be extracted from Figures 3 and 4 where we show the second order structure function calculated over the velocity image, and the corresponding kinetic energy spectrum $E(k)$, respectively, as a function of the wave number $k=2\pi/r$.

It is clear from both Figures that there is a well-developed turbulent spectrum over a wide range of scale lengths, though it does not correspond to pure Kolmogorov turbulence. The spectrum can be divided into two power law regions of different spectral indices, which does not resemble, in principle, the three-dimensional turbulent case. Main features of the spectrum are the lack of a low-$r$ cutoff, the presence of a knee at scales of $r \approx 10$ pc and a large scale turn over beyond $r \approx 60$ pc. The upper wavenumber-cutoff of the spectrum is located at $r \approx 1$ pc in Figures 3-4 (i.e., the size of the pixel), and so the viscous dissipation scale inside the cloud is certainly smaller than $3.6$ pc, the resolution limit of the image because of seeing effects. If energy is, as usual, cascading down along the spectrum, then there seems to be a priori two inertial ranges. A source would be located at intermediate scales, say $\approx 10$ pc, and another one at the largest observed scales (beyond $\approx 60$ pc). Scales of $\approx 10$ pc may correspond to energy injection by the terminal winds of OB cavities in the many regions of recent stellar formation or WR stars. But what about the less well defined power law extending up to the largest scales of the cloud ? What can be releasing energy into the system at so large scales ?

One possible energy source is gravitational interaction with external systems. In this context, Larson (1979) and Fleck (1981,1983) suggested that the energy associated with the turbulence observed in giant HII regions may be attributed to the differential rotation of their host galaxies. These authors proposed that the ionizing photon flux, originated at the border of molecular clouds, diffuses through a large volume of rarefied inter-cloud gas forming a giant HII region. Due to its large volume, the ionized cloud undergoes the effects of the differential rotation of the galaxy, which powers the turbulence at the largest scales and the subsequent cascade down to the dissipation region. The problem with this view is that giant HII regions with high dispersion velocities are also observed in galaxies of low rotation velocity. An example of the latter case is NGC 4449, an irregular galaxy of negligible rotation, where several giant HII regions with a high supersonic dispersion velocity are observed (Gallagher and Hunter 1983).

Furthermore, a problem is also posed by the mere existence of two regimes. As pointed out by Münch (1958), in discussing von Hoerner's (1955) work on the turbulence in Orion Nebula, the effective depth of line formation must be considered in the analysis of the observations in terms of some hydrodynamic model. Following Münch (1958), and assuming a constant gas density over the cloud's volume, $\rho(s) =$ const, and optical density $\tau(s)=Ks$, where s is the coordinate along the line of sight, the second order structure function can be written:

$$F_2(\Lambda) = K^2 \times \int_0^\infty \int_0^\infty e^{-K(s_1+s_2)} \cdot \left\{ \left\langle \left[ v_\alpha(s_1) - v_\beta(s_2) \right]^2 - \left[ v_\alpha(s_1) - v_\alpha(s_2) \right]^2 \right\rangle \right\} \cdot ds_1 ds_2 \qquad (6)$$

where $\alpha$ and $\beta$ designate two points on the two-dimensional image of the cloud separated a projected distance $\Lambda$. Under the assumption of homogeneous and isotropic Kolmogorov turbulence, the average squared velocity differences in the integrand of eq. (6) can be written as:

$$\left\langle \left[ v_\alpha(s_1) - v_\alpha(s_2) \right]^2 \right\rangle = C^2 \cdot \left| s_1 - s_2 \right|^{2/3}$$

(7)

$$\left\langle \left[ v_\alpha(s_1) - v_\beta(s_2) \right]^2 \right\rangle = C^2 \cdot \left[ \Lambda^2 + \left( s_1 - s_2 \right)^2 \right]^{1/3}$$

The resultant second order structure function tends to different power laws for both, small and large scales:

$$F_2(\Lambda) \propto \Lambda^{2/3} \qquad / \quad \Lambda >> K^{-1}$$

(8)

$$F_2(\Lambda) \propto \Lambda^{5/3} \qquad / \quad \Lambda << K^{-1}$$

Solutions of eq. (6) for various effective depths for line formation, $K^{-1} = 1, 5$ and $10$ pc, are shown in Figure 5.a. It can be seen that, even if the asymptotic values of the slope are similar, the transition between both power laws is much more gradual for the theoretical curve. Indeed, while the observed structure function needs less than half a decade to make the transition, the theoretical curve requires two decades to do the same. Therefore, a Kolmogorov like turbulence spectrum seems unlikely.

Another, more interesting possibility, which does not suffer from the previous shortcomings, is a *cascading-up* or *inverse transfer* of energy. That is, a part of the energy injected into the system at scales of the order of *10-20* pc goes into the formation of the larger structures.

The Kolmogorov viewpoint of energy cascade from large to small scale has often been opposed with the experimental evidence that large scales pair and amalgamate, leading to the formation of larger structures. In spite of the evidences, these facts are almost always neglected in the literature related with turbulence in giant H II regions (and in astronomy in general); consider as a typical example the asseveration "Turbulence means energy transfer from large eddies to smaller ones until it dissipates …" taken from a classical paper on the subject (Roy et al. 1986). However, there does not seem to be any contradiction at all between both mechanisms which certainly occur simultaneously, at least in a two-dimensional context. Two-dimensional turbulence has both kinetic energy and mean square vorticity as inviscid constants of motion, which allows the coexistence of two inertial ranges (e.g., Kraichnan 1967; Kraichnan and Montgomery 1980). If energy is injected at a constant rate in the neighborhood of wavenumber $k_i$, and the Reynolds number is large, a quasi-steady-state develops with $E(k) \propto \varepsilon^{2/3} k^{-5/3}$ for $k < k_i$, and $E(k) \propto \eta^{2/3} k^{-3}$ for $k > k_i$, where $\varepsilon$ is the rate of cascade of kinetic energy per unit mass, $\eta$ is the rate of cascade of enstrophy (mean-square vorticity) and the kinetic energy per unit mass is $\int_0^\infty E(k)dk$. Inside the "*-5/3*" range, kinetic energy is cascading backwards from higher to lower wave numbers $k$, and the vorticity flow is zero. On the other hand, inside the "*-3*" range, just the opposite takes place with an upward cascading vorticity flow and a zero-energy flow. Therefore, in a practical situation for example, the large scales of the fluid may be quasi-two-dimensional (in the mixing layer, for instance) and obey the two dimensional vorticity conservation constraint which implies strong inverse transfers of energy. On the other hand, they will simultaneously degenerate, through successive instabilities due to the three dimensional perturbations they are submitted to, towards small scale Kolmogorov fully developed three-dimensional turbulence which will dissipate the kinetic energy of the large scales of the mean flow.

The development of these concepts was motivated fundamentally by meteorological problems, where they are of practical application. For example, this seems to be the case in the Earth's atmosphere mesoscale (1 km - $10^2$ km) and the ocean (Lesieur 1987). In the atmosphere, small scale 3D turbulence (produced by the breaking of waves behind the mountains or by convective storms), may find afterwards imbedded into a stable stratified inversion density profile. A conversion of 3D into 2D turbulence has been proposed (Riley, Metcalfe and Weissman 1981) as a way of reorganizing the turbulence under the action of gravity, leading to the formation of *pancake shaped* collapsed eddies with 2D dynamics (Gage 1979; Lilly 1983), and feeding an inverse horizontal $k^{-5/3}$ energy cascade responsible for the atmospheric mesoscale energy spectrum extending up to several hundred kilometers. However, examples span from toroidal laboratory plasma confinement experiments (Zweben et al. 1979) and reversed field pinch experiments (Robinson and Rusbridge 1971) to fluctuations and large scale shares in the solar wind (e.g., Barnes 1979; Matthaeus and Goldstein 1982, Roberts et al. 1992) where the magnetic field provides the necessary anisotropy.

Confirmations of the amalgamation processes of two-dimensional eddies and of the development of the downward and upward cascade have been given, for example, by numerical simulations of Staquet (1985) and Frisch and Sulem (1984), respectively. In the first case, an inflectional velocity profile in the mixing layer develops into Kelvin-Helmholtz vortices that amalgamate into larger structures. In the second case, energy is injected in a narrow wavelength band to a fluid initially at rest, and the formation and evolution of the direct enstrophy cascade and inverse energy cascade at both sides of this range is clearly seen as predicted by the theory. Furthermore, the development of the inverse cascade can be viewed as resulting from the coalescence of eddies with comparable sizes (see also Herring and Williams 1984).

We also note that the values of the spectral indices obtained from our calculations (cf. Figure 4) are very nearly those predicted by the theory, i.e., +6.6 % and -15 % for the kinetic energy and enstrophy cascade, respectively.

The effect of the optical depth of line formation on the observed spectrum for two-dimensional turbulence can be evaluated using eq. (6) and the following expressions for the average velocity differences:

$$\left\langle \left[ v_\alpha(s_1) - v_\alpha(s_2) \right]^2 \right\rangle = C^2 \cdot \left| s_1 - s_2 \right|^{2/3}$$

$$\left\langle \left[ v_\alpha(s_1) - v_\beta(s_2) \right]^2 \right\rangle = C^2 \cdot \left[ \Lambda^2 + \left( s_1 - s_2 \right)^2 \right]^{1/3}$$

(9)

for $r > r_{inj}$ and,

$$\left\langle \left[ v_\alpha(s_1) - v_\alpha(s_2) \right]^2 \right\rangle = \frac{C^2}{r_{inj}^{4/3}} \cdot \left| s_1 - s_2 \right|^2$$

$$\left\langle \left[ v_\alpha(s_1) - v_\beta(s_2) \right]^2 \right\rangle = \frac{C^2}{r_{inj}^{4/3}} \cdot \left[ \Lambda^2 + \left( s_1 - s_2 \right)^2 \right]$$

(10)

for $r < r_{inj}$, where $r_{inj}=2\pi/k_{inj}$ is the scale at which energy injection occurs. The results for $r_{inj}=10$ pc, and $K^{-1}=1,5,$ $10$ and $20$ pc are shown in Figure 5.b. It can be seen that the rapid transition between both power laws in the theoretical curves closely resemble the shape of the observed structure function. Furthermore, in spite of the fact that the power law indexes are not exactly the same for theoretical and observed curves, the relationship between them is the same for each curve (e.g., a 5° clockwise rotation of the $K^{-1}=20$ pc curve perfectly matches the structure function of NGC 604). The position of the turnover determines the injection scale, $r_{inj} \approx 10$ pc, while the inverse cascade region of the spectrum favors rather large optical depths in the cloud, $K^{-1} \approx 20$ pc.

The velocities involved in Figure 3 are subsonic from the smallest resolved scale up to probably $r \approx 40$ pc. Supersonic velocity differences appear only between points separated by larger distances, at scales where only few eddies exist. This is remarkable because most previous works on turbulence in HII regions deals with supersonic motions (e.g., Tenorio-Tagle 1979; Bodenheimer et al. 1979; Terlevich and Melnick 1981; Fleck 1983; Roy and Joncas 1985; Roy et al. 1986; Melnick et al. 1987, Muñoz-Tuñón et al. 1996, Yang et al. 1996). Note, however, that they use line widths, σ, as diagnostic tools. Hence, their conclusions refer to the smallest, unresolved, scales which are actually dominated by supersonic motions. Inside each gas column element (pixel) there are probably many high velocity components, that can be associated with small shell fragments and condensations or high speed low mass stars, which widen the line profiles while participating also of the global, large-scale, mostly subsonic movements of the cloud. In the present work, we fit a single Gaussian to the line profile inside each pixel and use the observed shift of the center of the line, as we are interested on the average movement of the column element and not on its internal velocity structure. Therefore, there are probably different hydrodynamic phenomena at play on different scales: a supersonic one, dominant at small, unresolved scales and responsible for the observed line widths in giant H II regions (and the existent σ-diameter and σ-luminosity relations - Melnick et al. 1987), and a mainly subsonic hydrodynamic turbulence responsible for their larger scale kinematics.

Although a single Gaussian fit is used for the definition of the average gas velocity in each pixel, line profiles have multiple peaks over some regions of the cloud (see, Muñoz-Tuñon et al 1996 and Yang et al 1996). Moreover, systematic differences seem to exist between profiles in the brighter parts of the nebula (where they appear single peaked) and the remaining regions. To test our assumption that a single Gaussian fit is representative of the average motion of the gas in each pixel to the effects of turbulence analysis, we have repeated our calculations of the correlation function $F_2$ for another three regions of the cloud, labeled as frames 1, 2 and 3 in Figure 1.a, and compared this results with the correlation function for frame 0. In Figure 6 we plot the new correlation functions against $F_2^{[frame\ 0]}(r)$. It can be seen that, despite quantitative differences, the qualitative agreement is remarkably good, validating our assumption and showing that our conclusions are position independent inside the cloud.

If the turbulence in NGC 604 is actually two-dimensional, and an inverse energy cascade is taking place, the main source of energy of the "-5/3" regime lies at intermediate scales inside the cloud and not outside as the differential rotation viewpoint claims. The flux of kinetic energy per unit mass, $\varepsilon$, inside the inverse cascade can be estimated in the following way. Consider a two-dimensional eddy (a cylinder) of radius $l$, thickness $h$, angular velocity $\omega$, and tangential velocity $v = \omega \times l$. The kinetic energy of the eddy is:

$$E = \frac{1}{2} I \omega^2 = \frac{1}{2} \left( \frac{M l^2}{2} \right) \omega^2 = \frac{1}{2} \left( \frac{M l^2}{2} \right) \left( \frac{v}{l} \right)^2 = \frac{M v^2}{4} \qquad (11)$$

Furthermore, assuming that the eddy transfers a substantial fraction of its energy during a revolution time, i.e., $\Delta t \approx 2\pi l / v$,

$$\varepsilon \approx \frac{E}{M} \cdot \frac{1}{\Delta t} = \frac{v^2}{4} \cdot \frac{v}{2\pi l} = \left( \frac{1}{8\pi} \right) \cdot \frac{v^3}{l} \quad . \qquad (12)$$

where $v$ and $l$ can be obtained from the structure function given in Figure 3 at any point inside the kinetic energy inertial range. This gives $\varepsilon \approx 3 \times 10^{-4}$ erg g$^{-1}$ sec$^{-1}$. On the other hand, the mass of gas (HI + HII) inside the cloud is $M_g \approx 3 \times 10^6\ M_\odot$ (Wright 1971; Israel et al. 1982), while the mass in ionizing (OB) stars is $M_* \approx 1.5 \times 10^4\ M_\odot$ (Kennicutt, 1984). If we call $L_T$ the power of the sources of the turbulence, then $\varepsilon = L_T / M_g$. Writing the fraction of gas in ionizing stars as $\zeta = M_* / M_g \approx 8 \times 10^{-3}$, the power of the sources of kinetic energy per unit mass of ionizing star is:

$$\frac{L_T}{M_*} = \frac{\varepsilon}{\zeta} \approx 4 \times 10^{-2} \qquad (13)$$

If we consider that the sources are common OB stars with a typical $L_{bol} / M \approx 10^2$, then the fraction of their bolometric luminosity needed to power the turbulence observed in the inertial range is:

$$\frac{L_T}{L_{bol}} = \frac{L_T}{M_*} \times \left(\frac{M}{L_{bol}}\right)_{OB} \approx 4 \times 10^{-4} \qquad (14)$$

Therefore, just a few WR stars would be enough to sustain the observed turbulence. Following D'Odorico and Rosa (1981), approximately 50 WR stars should be present in the region. Therefore, $\zeta_{WR} = M_{WR} / M_g \approx 3 \times 10^{-4}$, where $M_{WR}$ is the mass of the 50 WR and,

$$\frac{L_T}{M_{WR}} = \frac{\varepsilon}{\zeta} \approx 1 \quad , \qquad (15)$$

giving for the fraction of the bolometric luminosity going into turbulent kinetic energy,

$$\left(\frac{L_T}{L_{bol}}\right)_{WR} = \frac{L_T}{M_{WR}} \times \left(\frac{M}{L_{bol}}\right)_{WR} \approx 10^{-5} \qquad (16)$$

which is of the same order of magnitude as $L_w / L_{bol}$ (where $L_w$ is the kinetic energy power carried by the stellar wind) for a WR star. Consequently, a few tens of WR stars (say, 50) are enough to power the inverse kinetic energy cascade observed.

On the other hand, a crude estimate of the radius of the ionization front of a generic massive OB star is (Franco et al. 1994):

$$R_{OB} \approx \left(\frac{3F_*}{4\pi n^2 \alpha}\right)^{1/3} \approx 10 \cdot \left(\frac{F_*}{4 \times 10^{48}}\right)^{1/3} n_{10} \quad pc \qquad (17)$$

where $F_*$ is the Lyman-continuum UV flux in units of sec$^{-1}$, and $n_{10}$ is the density in units of 10 cm$^{-3}$; which is just about the right scale for mechanical energy injection. Furthermore, the equilibrium radius of wind bubbles can be estimated as:

$$R_W \approx 2 \times \left(\frac{\dot{M}_{-5} V_8}{P_{-10}}\right)^{1/2} \quad pc \approx 2 \times \left(\frac{L_{36}}{V_8 P_{-10}}\right)^{1/2} \quad pc \qquad (18)$$

where $L_{36}$ and $\dot{M}_{-5}$ are the wind luminosity and mass loss in units of $10^{36}$ erg/sec and $10^{-5}$ $M_\odot$/yr respectively, $V_8$ is the wind velocity in units of $10^8$ cm/sec and $P_{-10}$ is the external pressure in $10^{-10}$ dyn/cm$^2$. Therefore, either single stars or associations of at most few stars, should be able to blow wind cavities with the right sizes.

Nevertheless, even if the energy requirements are numerically modest and the proper scales can occur naturally in the system, it is not clear whether the energy can be released coherently in the relatively large scales and narrow band required, $r_{inj} \approx 8$-10 pc. Much the same applies to other interesting mechanisms that have been proposed in the literature such as the share effects triggered by Kelvim-Helmholtz instabilities during the bursting and formation of individual H II regions (Blake 1972, Bodenheiner et al. 1979, Roy and Joncas 1985) or by champagne flows (Tenorio-Tagle 1979).

The low-wavenumber turnover of the structure function, carries information about the age of the turbulence, and evolves according to $k_E \propto \varepsilon^{1/2} \times \tau^{-3/2}$ where $\tau$ is the age of the turbulence. Therefore, the inverse cascade is quasi-stationary and continuously propagates to ever lower wavenumbers (Lesieur, 1987). However, the system we are studying is finite and, in that case, the energy accumulates at the lowest possible wavenumber at which the borders are reached (e.g., Kraichnan 1967; Kraichnan and Montgomery 1980). Such an effect is not observed in Figure 4, and so we conclude that the turbulence in NGC 604 is relatively young. This is another point in favor of the young massive stars as the engine feeding the turbulent spectrum, as the stellar formation process is recent ($\approx 4$ Myr) in NGC 604 (there is only one supernova remnant known in the region - Benvenuti et al., 1979). Inverting the reasoning, we can use the age of the evolution process and calculated energy flux per unit mass, to determine the proportionality constant in the expression for $k_E$:

$$\alpha = \frac{k_E}{\varepsilon^{1/2} \cdot \tau^{-3/2}} \approx 3 \times 10^2 \quad [cgs] \tag{19}$$

where $k_E \approx 2\pi/60$ pc$^{-1}$ was used. Therefore, the largest scale reached by the inverse energy cascade increases with time as:

$$\left(\frac{l}{1pc}\right) \approx 7.5 \times \left(\frac{t}{1Myr}\right)^{3/2} \tag{20}$$

and should extend up to the borders of the region during the next $\approx 10^6$ yr.

## 4. Conclusions

We show that the autocorrelation and structure functions have a considerable potential for the study of the dynamics of intergalactic HII regions when high resolution velocity images are available. In particular, by applying them to the study of NGC 604, we show that:

1) There are evidences of considerable structure in the velocity field projected along the line of sight. These include the probable presence of eddies ranging from scales of 10 pc to the largest scales observed in the cloud, and a rotation of the cloud as a whole with a period of 85 Myr.

2) A Kolmogorov like three-dimensional turbulence spectrum, though not impossible, seems unlikely.

3) A two-dimensional turbulent spectrum, powered at intermediate scales $r_{inj} \approx 10$ pc, is consistent with the observations if the optical depth for line formation is $K^{-1} \approx 20$ pc. This energy spectrum consists of a double cascade: a direct enstrophy cascade and an inverse energy cascade; the observed spectral indices being very nearly those predicted by the theory (differences of approximately -15% and +6.6% respectively).

4) A few tens of WR stars would be enough to feed the necessary kinetic energy through their heavy and fast mass losses in the right length scales.

5) The presence of a smooth low-wavenumber turnover indicates that the turbulence is relatively young and the inverse energy cascade had not enough time to extend up to the largest scales of the region. This gives another element in favor of the turbulence being powered from the inside by the recent process of stellar formation ($\approx$ few x $10^6$ yr old).

**ACKNOWLEDGEMENTS**

The authors thank the Brazilian agency FAPESP for its support. VJP and RO would like to thank the Brazilian agency CNPq for partial support. GAMT benefited from discussions with G. Tenorio-Tagle, R. J. Terlevich, C. Muñoz-Tuñon and J. Franco.

**Figure Captions**

**Figure 1: (a)** Hα peak intensity map and **(b)** radial velocity map of NGC 604, obtained by two dimensional high spatial resolution Fabry-Perot observations with TAURUS II at the 4.2 m William Herschel Telescope at the observatory of Roque de los Muchachos, La Palma, Spain. The results presented in Figures 2-5 are based on the analysis of the central frame in (b), also shown as frame **0** in (a). Frames **1**, **2** and **3** in (a) are used exclusively in the calculations of Figure 6.

**Figure 2:** Correlation function $F_1(r)$. **thick line:** auto-correlation function, eq. (1), calculated over the central portion of the radial velocity image (frame in Figure 1.b); **dotted line:** auto-correlation function minus the linear fit to the interval $r \in (50, 80)$ pc; **insert:** auto-correlation function minus the linear fit to the interval $r \in (1, 40)$ pc. See the text for details.

**Figure 3:** Structure function evaluated over the radial velocity image. The bars are a measure of the dispersion of $\langle [v(\underline{x}) - v(\underline{x} + \underline{r})]^2 \rangle$.

**Figure 4:** Main result of this work. Kinetic energy power spectrum of the turbulence in NGC 604, evaluated from the structure function. Note the double power law cascade at both sides of the knee at scales of $\approx 10$ pc. This kind of spectrum is typical of two-dimensional turbulence, where energy is being injected into the system at scales between approximately 10 and 20 pc (see insert). There is an inverse flux of kinetic energy (at zero enstrophy flux) towards low wavenumbers, while enstrophy cascades up to higher wavenumbers at zero energy flux. The position of the turn over of the power spectrum at the lowest wavenumbers is not stationary, and carries information about the time of the onset of the energy source. The theoretical values of the power law indices (see insert) are "-5/3" and "-3"

for the kinetic energy and enstrophy cascades, respectively, while the observed values are "-1.78" and "-2.55".

**Figure 5:** **(a)** second order structure function for a Kolmogorov turbulence when optical depth effects are taken into account (see, Münch 1958). The theoretical curves are for $K^{-1} = 1, 5, 10$ and $20$ pc.**(b)** Same as (a) but for two-dimensional turbulence. The curves are for $r_{inj} = 10$ pc and $K^{-1} = 1, 5, 10$ and $20$ pc. The best fit is provided by the two-dimensional turbulence structure function for large optical depth ($K^{-1} = 20$ pc), which can reproduce the rapid change in slope seen in the observed structure function as well as its shape.

**Figure 6:** comparison between the correlation functions evaluated inside frames **1**, **2** and **3** (vertical axis) and the correlation function inside frame **0** (horizontal axis). It can be seen that, although quantitative differences exist, the qualitative behavior of the correlation function is independent of position inside the cloud. As line profiles are different in each region, this figure shows the validity of the single Gaussian fitting approximation for the analysis of the large scale turbulence in the cloud.

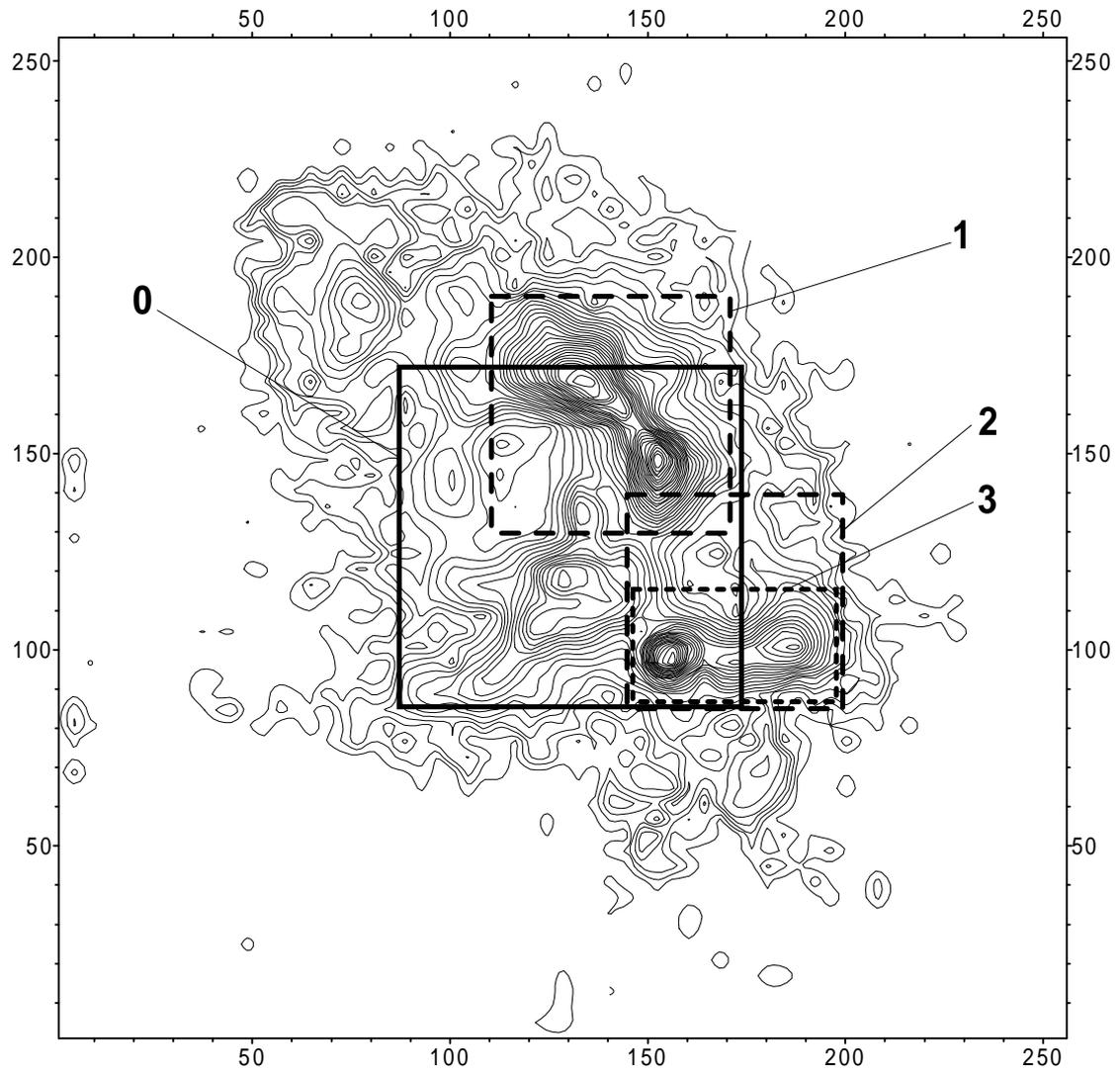

Figure 1.a

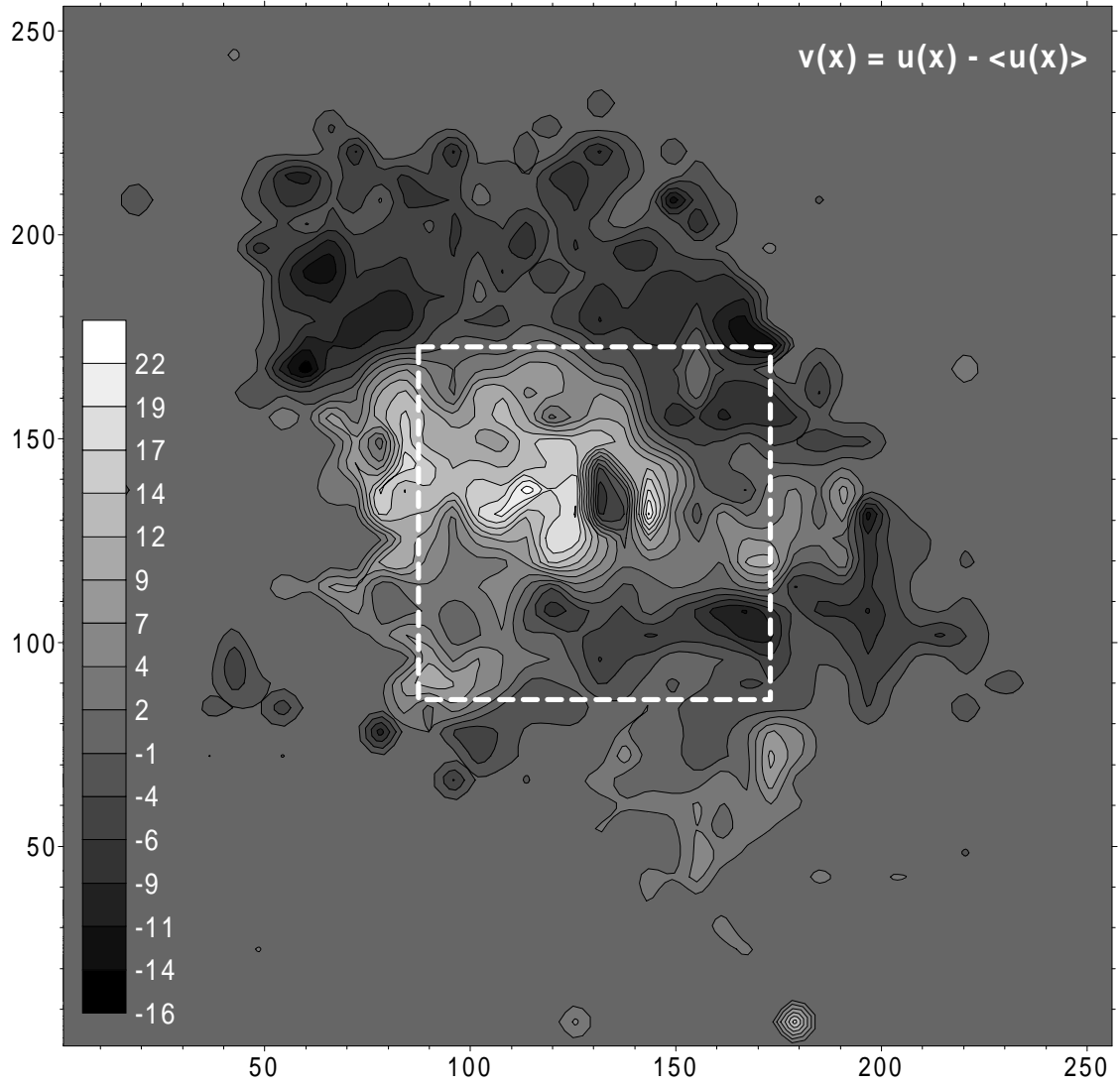

Figure 1.b

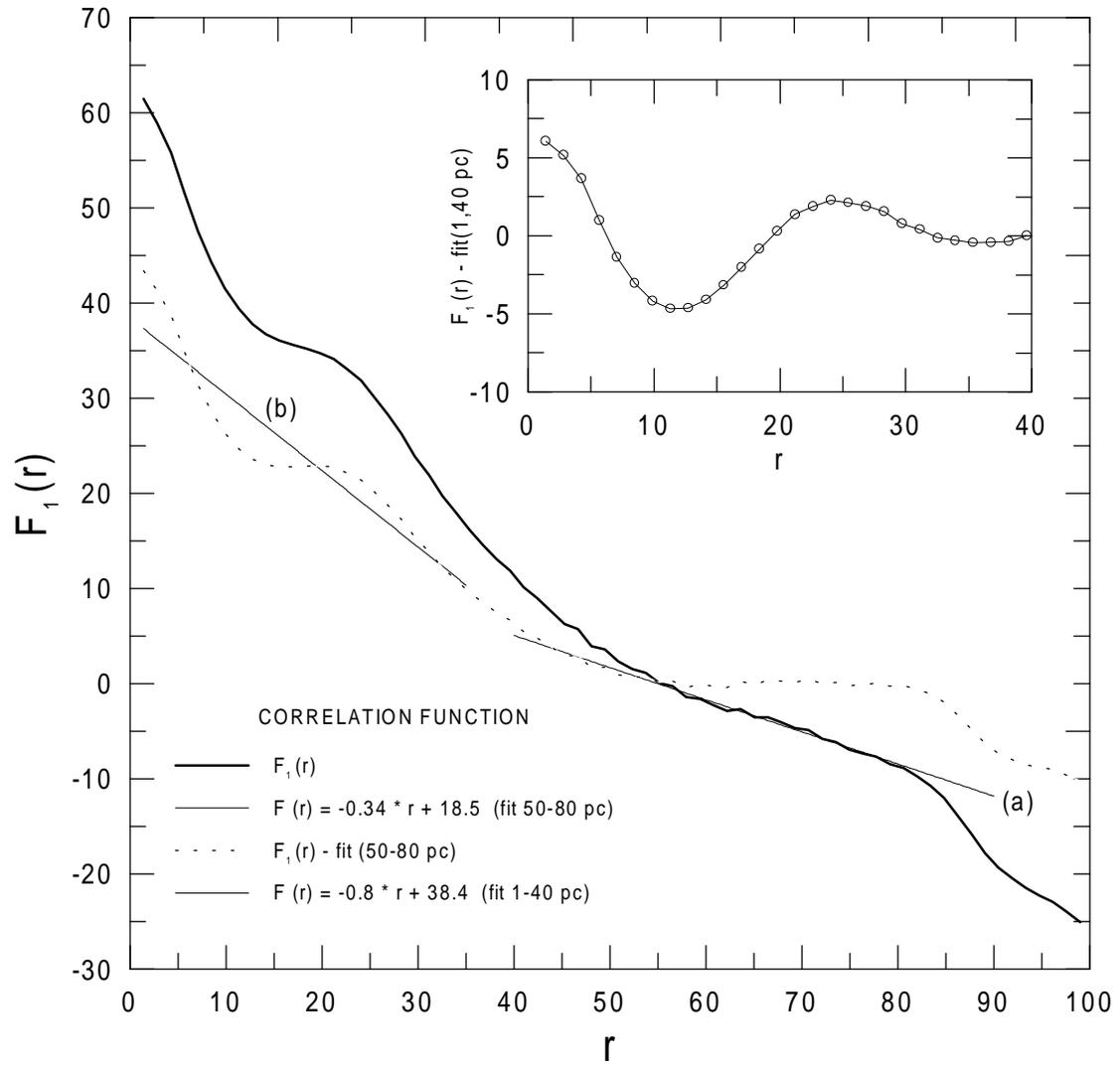

**Figure 2**

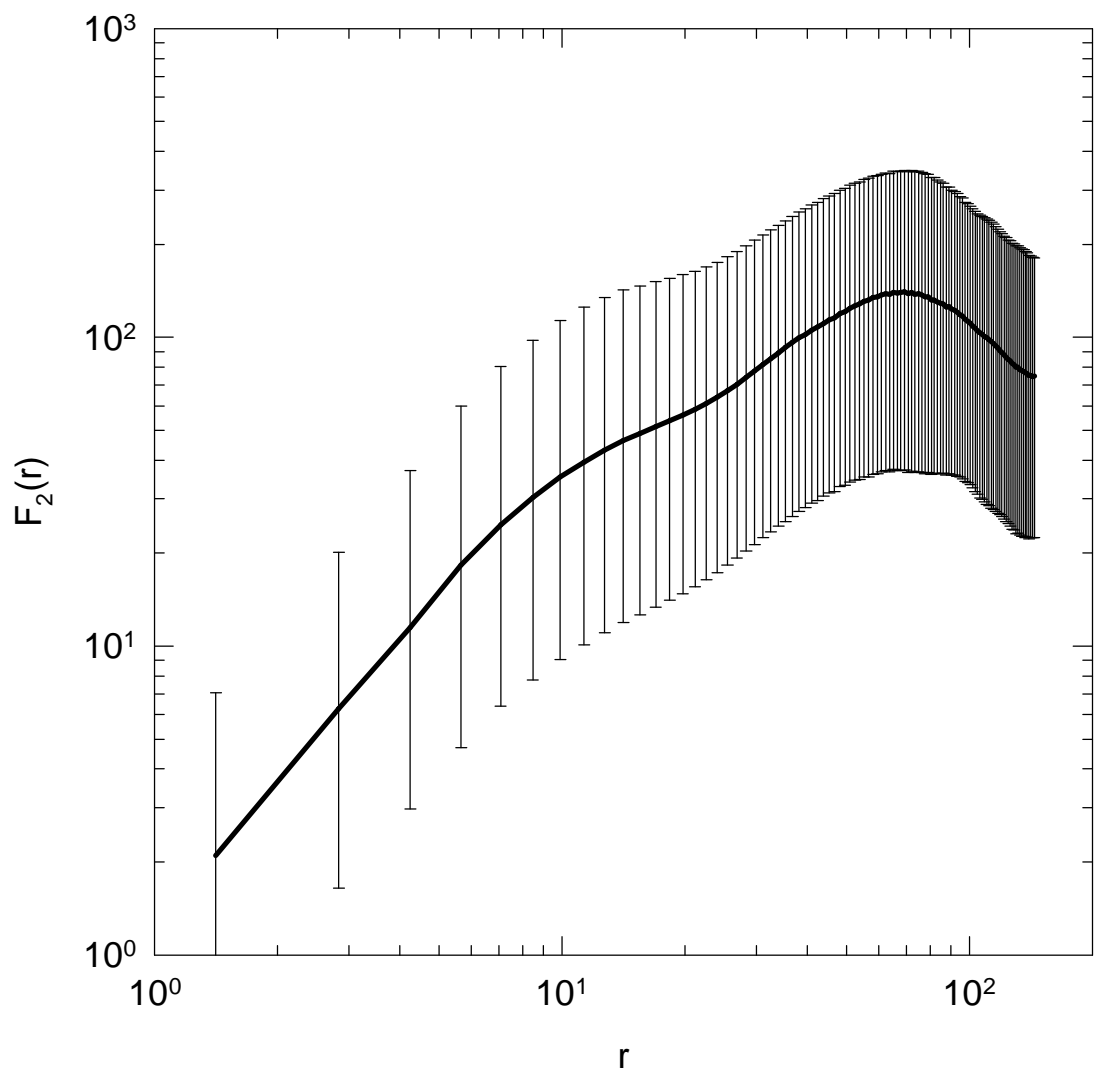

**Figure 3**

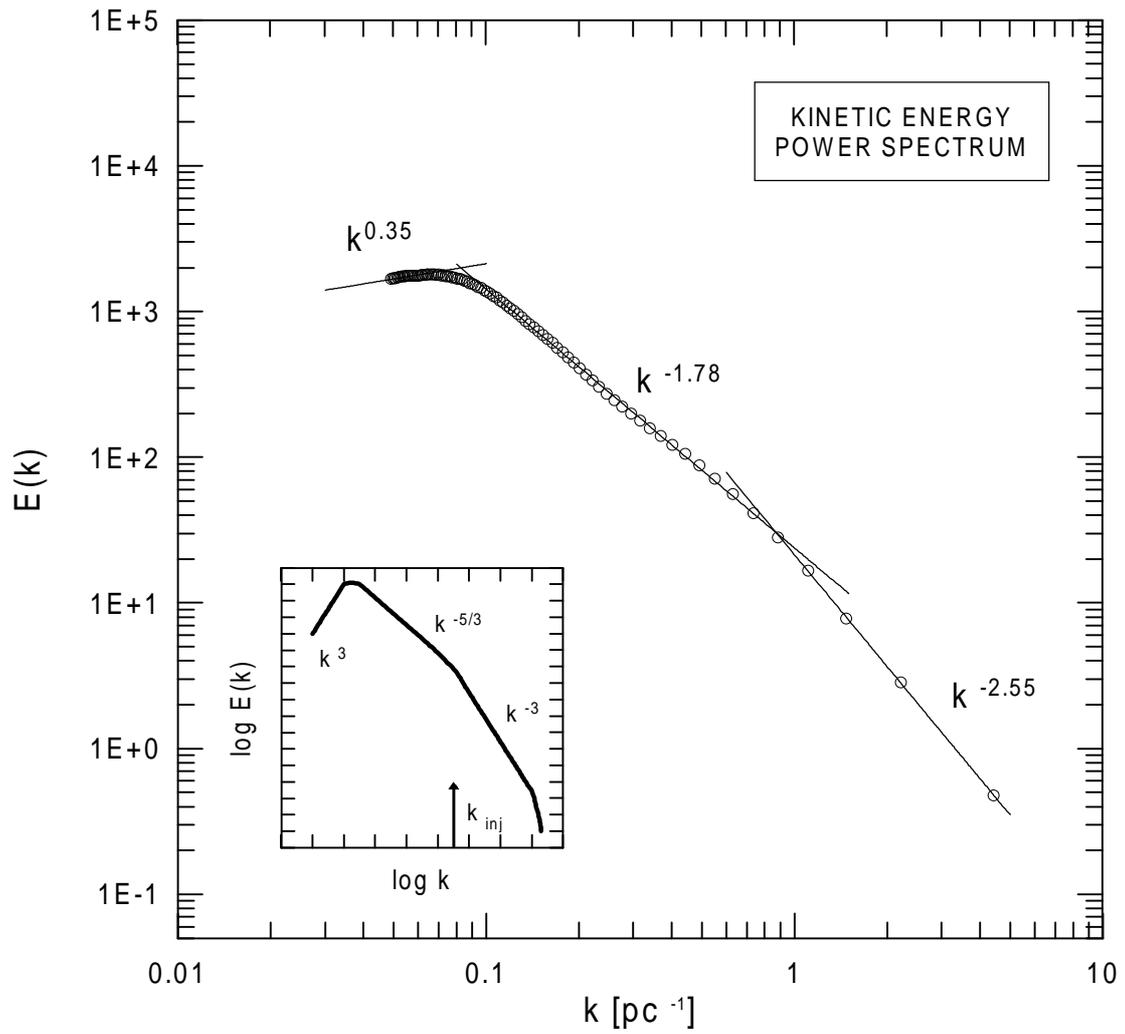

Figure 4

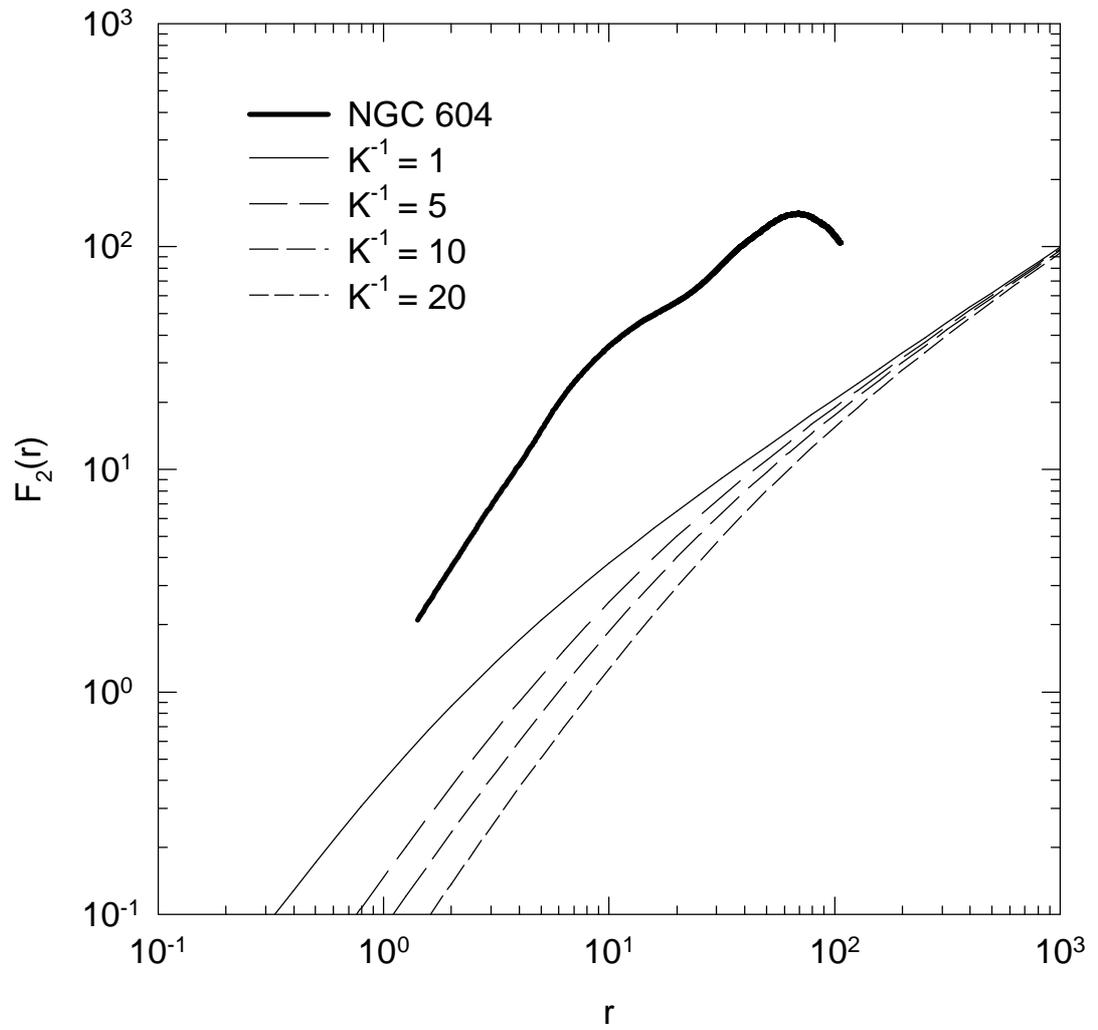

**Figure 5.a**

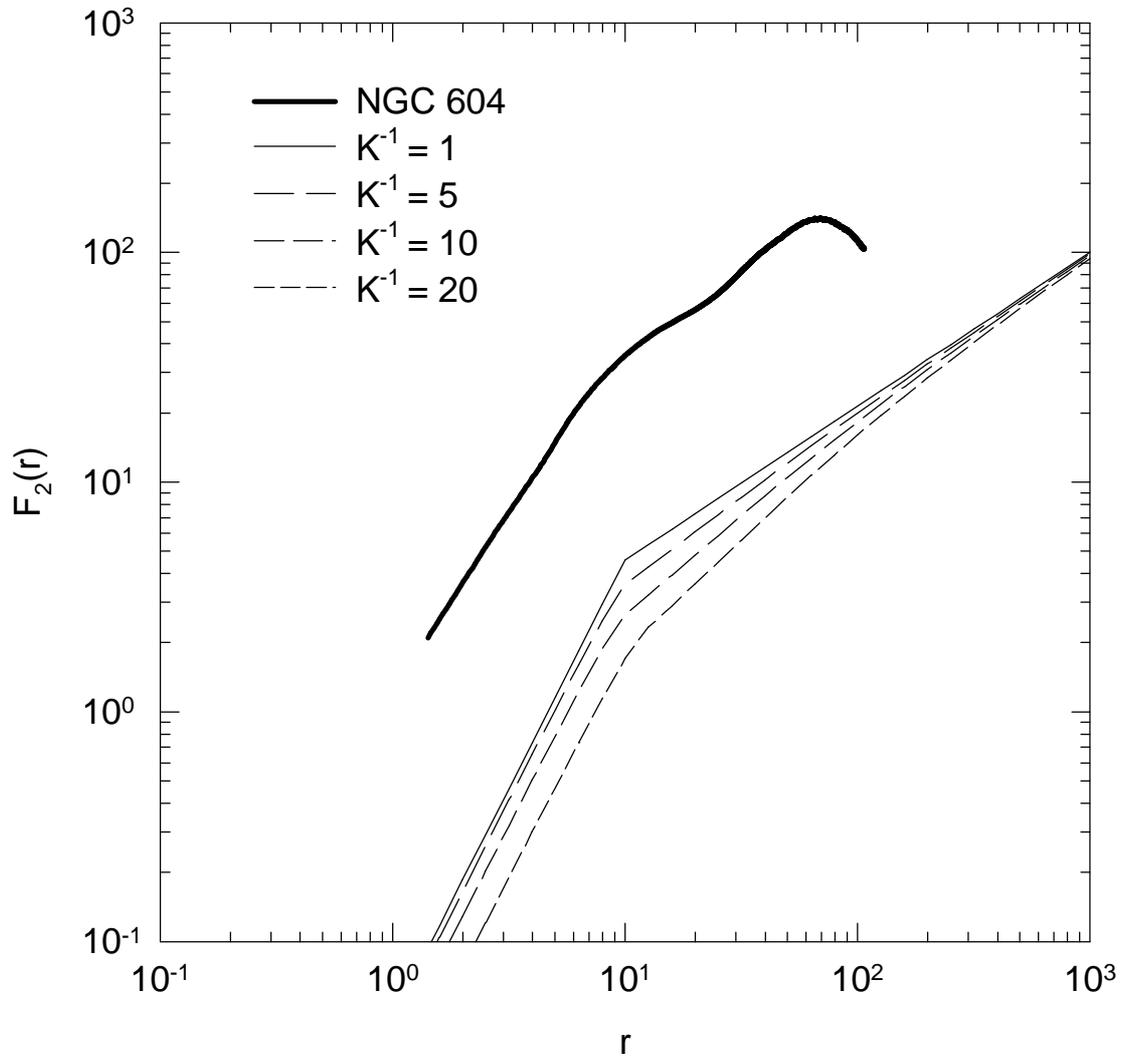

**Figure 5.b**

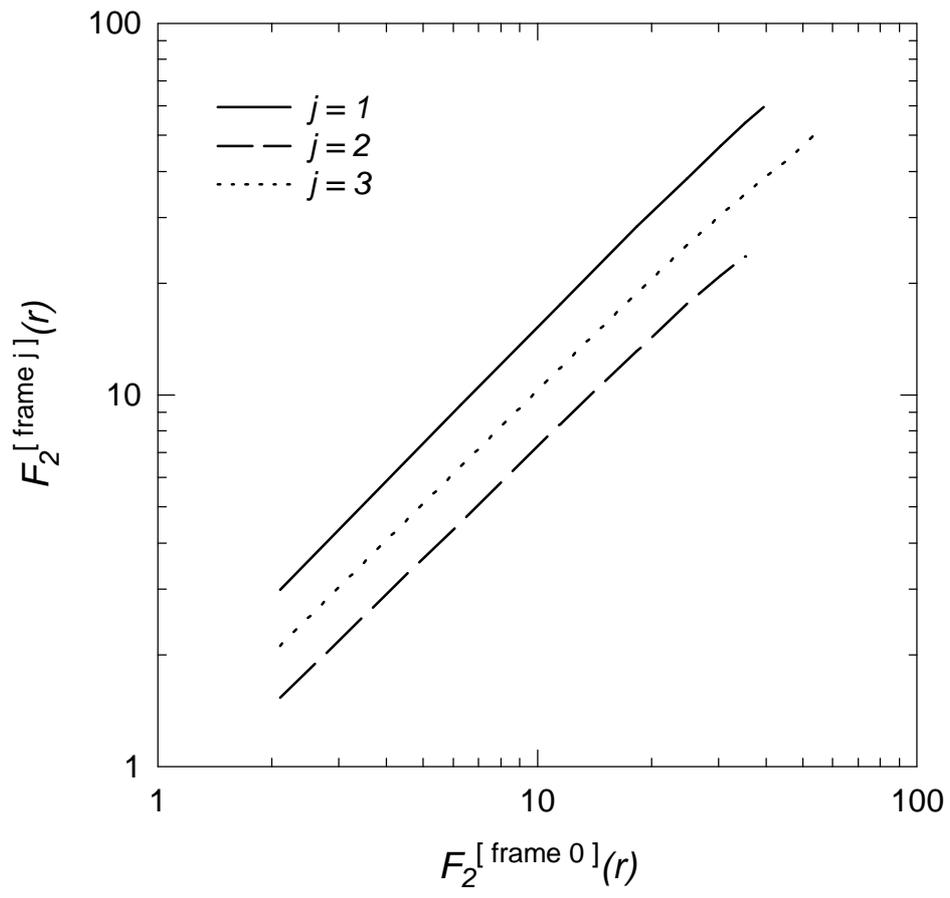

**Figure 6**